\documentclass[aps,prb,twocolumn,amsmath,amssymb,showpacs,letterpaper]{revtex4-1}

\usepackage[pdftex,colorlinks, linkcolor={blue}, citecolor={blue}]{hyperref}
\usepackage{graphicx}
\usepackage{dcolumn}
\usepackage{bm}

\begin{document}

\title{Spectroscopic signatures of the Larkin-Ovchinnikov state in the conductance characteristics of a normal-metal/superconductor junction}

\author{Qinghong Cui$^{1,2}$, C.-R. Hu$^3$, J.Y.T. Wei$^{4,5}$, and Kun Yang$^2$}

\affiliation{$^1$Key Laboratory of Yunnan Higher Education Institutes for Advanced Functional and Low Dimensional Materials and College of Physics and Electronic Engineering, Qujing Normal University, Qujing, Yunnan 655011, China}
\affiliation{$^2$National High Magnetic Field Laboratory and Department of Physics, Florida State University, Tallahassee, Florida 32306, USA}
\affiliation{$^3$Department of Physics, Texas A\&M University, College Station, Texas 77843, USA}
\affiliation{$^4$Department of Physics, University of Toronto, 60 St. George St. Toronto, ON M5S1A7, Canada}
\affiliation{$^5$Canadian Institute for Advanced Research, Toronto, M5G1Z8, Canada}

\date{\today}

\begin{abstract}
Using a discrete-lattice approach, we calculate the conductance spectra between a normal metal and an $s$-wave Larkin-Ovchinnikov (LO) superconductor, with the junction interface oriented {\em along} the direction of the order-parameter (OP) modulation. The OP sign reversal across one single nodal line can induce a sizable number of zero-energy Andreev bound states around the nodal line, and a hybridized midgap-states band is formed amid a momentum-dependent gap as a result of the periodic array of nodal lines in the LO state. This band-in-gap structure and its anisotropic properties give rise to distinctive features in both the point-contact and tunneling spectra as compared with the BCS and Fulde-Ferrell cases. These spectroscopic features can serve as distinguishing signatures of the LO state.
\end{abstract}


\maketitle

\section{Introduction}

When a spin-singlet superconductor is subjected to a Zeeman magnetic or exchange field, the Fermi surfaces of spin-up and -down electrons can undergo energy splitting. If this pair-breaking field is sufficiently strong, the order parameter (OP) can become spatially periodic, as proposed by Fulde and Ferrell~\cite{FF} and by Larkin and Ovchinnikov~\cite{LO} independently. In the Fulde-Ferrell (FF) scenario, the pairing is between $({\bf k}+{\bf q}/2, \uparrow)$ and $(-{\bf k}+{\bf q}/2, \downarrow)$ electrons, which results in an order parameter of the form $\Delta_q \exp i{\bf q} \cdot {\bf x}$ with a winding phase factor, where ${\bf q}$ is the pairing momentum. In the Larkin and Ovchinnikov (LO) scenario, the OP is spatially modulated with periodic sign reversal, the simplest case being $2\Delta_q \cos {\bf q} \cdot {\bf x}$. Such pairing states are now collectively known as the Fulde-Ferrell-Larkin-Ovchinnikov (FFLO) state. This novel inhomogeneous superconducting state has attracted broad theoretical interest~\cite{machida, shimahara97, kun98, shimahara98, kun01, agterberg01, krawiec, dalidovich, mora, mizushima, agterberg08, cui, miyaki, quan, yokoyama, fujimoto} due to the experiments suggestive of its existence in various superconductors such as heavy fermion,~\cite{matsuda, pfleiderer, koutroulakis} organic and other superconductors,~\cite{singleton, organic} and its possible realization in cold-atom systems,~\cite{coldatom, liao} high-density quark matter, and nuclear matter.~\cite{casalbuoni} Although it is long believed that the FFLO state can only exist in unconventional superconductors, experimental indication of disordered FFLO phase was reported in a conventional superconductor recently.~\cite{loh} However, direct evidence for the periodic OP variation is still desirable.~\cite{kun00, bulaevskii, kun04, vorontsov, wang, ikeda}

To help identify the FFLO state unambiguously, we previously proposed using conductance spectroscopy of normal-metal/superconductor (N/S) junctions as an experimental probe, treating the FF state first as an illustrative case in that work.~\cite{FFcase} However, only the spectral characteristics of a momentum-dependent gap due to a single non-zero pairing momentum are discussed there. Here, we show that the periodic OP sign reversal of the LO state can lead to further distinctive features in both point-contact and tunneling conductance spectra when the junction interface is oriented perpendicular to the OP nodal lines. These features are the result of repeated intrinsic Andreev reflections around each nodal line in the bulk superconductor and can be used to distinguish the LO state from both the BCS and the FF states.

\begin{figure}
\includegraphics[width = 0.39\textwidth]{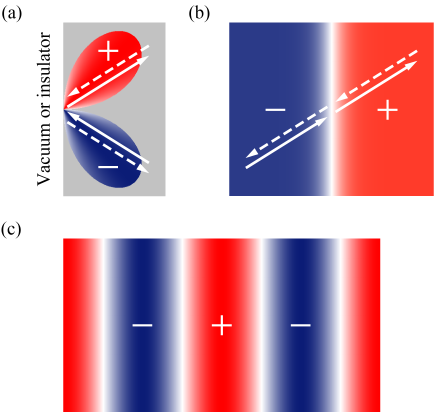}
\caption{\label{fig1} (Color online) Formation of the Andreev midgap states: (a) on the (110) surface of a $d_{x^2-y^2}$-wave superconductor; (b) about the nodal line of an $s$-wave superconductor with a $\tanh(x)$-like order parameter. In both cases, a quasi-electron (solid arrow) is retroreflected near the surface or the nodal line (white line) as a quasihole (dashed arrow) and vice versa. Essentially, zero-energy midgap states are formed near the interface due to the order-parameter sign reversal experienced by repeated Andreev reflections. (c) Schematic of the spatial variation of the OP of the LO state with periodic sign reversal. The Andreev midgap states, which form in the vicinities of the periodically spaced nodal lines, are coupled to become a hybridized midgap-states band.}
\end{figure}

When an electron from N is incident on S at an energy within the superconducting gap, it can enter S via a process known as Andreev reflection,~\cite{bruder, deutscher} whereby a hole of nearly equal momentum is {\em retroreflected} at the N/S interface and a Cooper pair emerges simultaneously in S. An important application of Andreev reflection is when the OP experienced by a quasiparticle changes sign upon a specular reflection at a barrier interface, such as in a finite-impedance normal-metal insulator superconductor (NIS) junction oriented along a nodal line of a $d_{x^2-y^2}$-wave superconductor [see Fig.~\ref{fig1}(a)]. Midgap surface states (MSS) of practically zero energy~\cite{hu} are formed near the interface due to repeated Andreev and specular reflections, in accordance with the Atiyah-Singer index theorem in topology.~\cite{ryu} A distinct manifestation of these MSS is the zero-bias conductance peak (ZBCP) observed in the N/S tunneling spectra on various unconventional superconductors,~\cite{deutscher, tanaka95, tunneling} sometimes with robust spectral height and sharpness.~\cite{wei} These MSS are also manifested in penetration-depth measurements.~\cite{walter, carrington}

Midgap quasiparticle states can also form about an isolated real-space nodal line of a $\tanh(x)$-like OP inside the superconductor as a result of repeated Andreev reflections alone, without involving specular reflection and thus the OP sign reversal in momentum space [see Fig.~\ref{fig1}(b)]. In the LO state, the intrinsic OP sign reversal over a periodic array of real-space nodal lines [see Fig.~\ref{fig1}(c)] will cause the formation of a hybridized midgap-states band (HMSB),~\cite{vorontsov, wang} which can not form in either the BCS or the FF states. The states in this HMSB, being anisotropic bulk quasiparticle states in nature, can facilitate the transmission of the incident electrons mainly along the nodal lines if only the nodal lines are intercepted by the junction interface.~\cite{transmission} These states will then reduce the probability for Andreev reflection at the interface. Consequently, novel features related directly to the periodically sign-reversing OP are manifested in the conductance spectra, depending on both the N/S junction orientation and impedance.

It should be noted that these bulk manifestations of the LO state are independent of the pairing symmetry since the formation of the {\em bulk} midgap states does not require momentum-space OP sign reversal. Therefore, $d$-wave pairing symmetry will not disrupt these bulk manifestations qualitatively, even though quantitative spectral differences are expected to appear. Furthermore, the {\em bulk} manifestations of the LO state can also be systematically distinguished from the {\em surface} manifestations of a $d$-wave OP arising from the MSS, i.e., the ZBCP in tunneling spectra,~\cite{hu} because a barrier layer is required to induce the MSS [see Fig.~\ref{fig1}(a)]. Therefore, the MSS can appear only in finite-impedance spectra as a midgap peak, but disappear in zero-impedance point-contact spectra. In contrast, the {\em bulk} HMSB states can appear in both finite- and zero-impedance spectra because they are essentially {\em bulk} quasiparticle states.

In order to demonstrate the mechanism of the HMSB in identifying the LO state, we focus on $s$-wave superconductors so that the MSS formed at the N/S interface barrier due to momentum-space OP sign reversal are excluded, and only the essential differences between the LO and FF states are illustrated. We present the calculated conductance spectra of the LO state with the N/S interface {\em parallel} to {\bf q} (hence perpendicular to the nodal lines) in a discrete-lattice model along with that of the FF state for comparison. A continuum model for the high-impedance junction where the interface is {\em normal} to {\bf q} (hence parallel to the nodal lines) has been given by Tanaka {\it et al}.~\cite{tanaka07}

This paper is organized as follows. In Sec.~II, we introduce the model and present the numerical results on the density of states (DOS) and conductance spectra of both the LO and the FF cases. After an analysis of the band structures of both the FF and the LO states, the manifestations of the HMSB in the conductance spectra, which occur in the LO state only, are discussed in Sec.~III. Concluding remarks are offered in Sec.~IV. Throughout this paper we consider zero temperature only in order to illustrate the essential physics.

\section{Model and numerical results}

\begin{figure*}
\includegraphics[width = 0.7\textwidth]{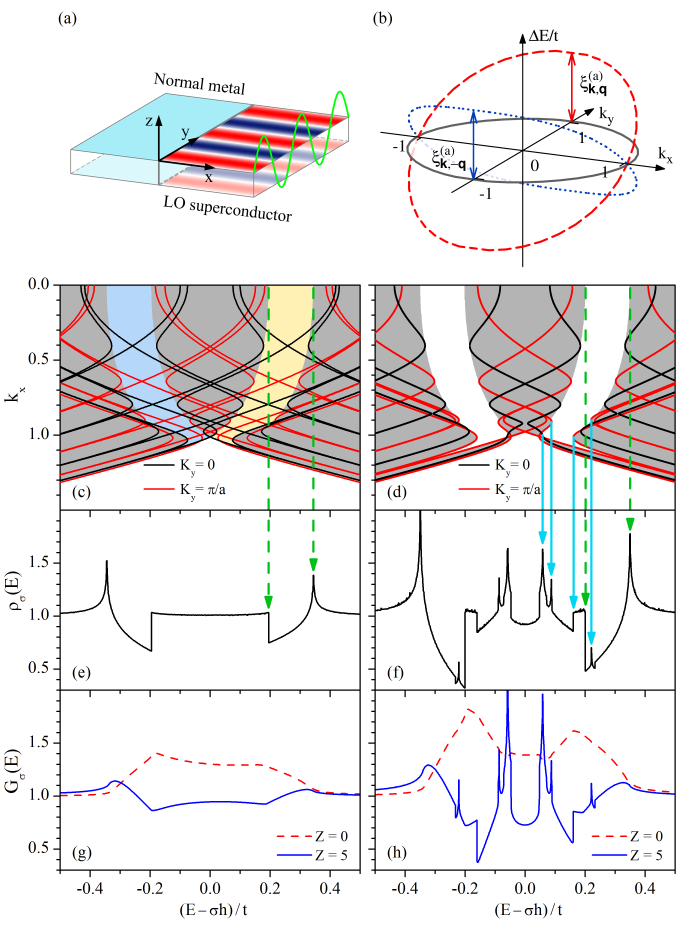}
\caption{\label{fig2} (Color online) Numerical simulations illustrating the effects of the hybridized midgap-states band on the DOS and N/S junction conductance spectra. 
(a) Schematic of an N/S junction for the $s$-wave LO case considered in our model, with the N/S interface ($x = 0$) parallel to the pairing momentum {\bf q}. The OP variation is assumed to be within the layers of the quasi-2D material, which are parallel to the $xy$ plane. (b) Trajectories of the gap centers of the FF states with pairing momenta of $+{\bf q}$ (red dashed line) and $-{\bf q}$ (blue dotted line), respectively, along with the Fermi surface of the normal state (solid line). (c), (d): The quasiparticle energy-momentum dispersions for $s$-wave FF (c) and LO (d) superconductors, respectively, with different crystal $y$ momenta $K_y$. The dispersion curves are symmetric about the true $x$ momentum $k_x = 0$. In (d), the gray shading marks out regions in $(k_x, E)$ space where states exist, while in (c), the gray shading is the overlap of the regions occupied by states of positive (blue shading) and negative (yellow shading) $k_y$. (e), (f): The normalized DOS $\rho_{\sigma}(E)$ for the FF (e) and LO (f) states in a bulk superconductor. (g), (h): The normalized conductances $G_{\sigma}(E)$ of the FF (g) and LO (h) states in the point-contact ($Z = 0$) and tunneling ($Z = 5$) limits. The green dashed (cyan solid) arrows indicate that the singularities in the DOS and $G_{\sigma}(E)$ spectra originate from states in the quasiparticle dispersion which have momenta nearly parallel (perpendicular) to the pairing momentum {\bf q}. Here $E$ is the quasiparticle energy, $\sigma$ is the spin index, $h$ the Zeeman energy, and $t$ the hopping integral. In this case, we take LO OP $\Delta_q = 0.075$ and the period length $a = 20$ for a better illustration of the fine features of the DOS and the junction conductance. See text for the values of other parameters used.}
\end{figure*}

Here, we consider a quasi-two-dimensional (2D) material, with the magnetic field applied parallel to the layers. We let the N/S junction interface be perpendicular to the layers. The {\bf q}-vector describing the one-dimensional OP variation is assumed to be in the layers and along the interface. We choose a coordinate system such that the $x$ axis is perpendicular to the N/S interface, and the $y$ axis is along {\bf q}. The layers of the material are therefore parallel to the $xy$ plane shown in Fig.~\ref{fig2}(a).  In this geometry, the orbital effect is very weak and can be neglected to a good approximation (especially if the sample thickness along $z$ is much smaller than the Josephson penetration depth of the sample). Thus, only conductance within the layers needs to be considered. For simplicity, we assume that the N side is similarly quasi-2D. We can then reduce the problem to a 2D problem.

We use a discrete square-lattice application of the Blonder-Tinkham-Klapwijk (BTK) theory.~\cite{BTK} The semi-infinite N and S regions are on the left ($x < 0$) and right ($x > 0$) sides of the interface, respectively. The barrier at the interface ($x = 0$) is modeled by a scattering potential $U_0 \delta_{x,0}$ with the barrier-strength parameter $Z = U_0/2t$, where $t$ is the hopping integral and is taken to be $1$ as the unit of energy. The quasiparticles of the system are described by the Bogoliubov-de Gennes (BdG) equations
\begin{equation} \label{eq:BdG}
\sum_{{\bf j}} \left(
\begin{array}{cc}
H_{{\bf ij}, \sigma} & \Delta_{{\bf ij}} \\
\Delta_{{\bf ji}}^{\ast} & -H_{{\bf ij}, \bar{\sigma}}^{\ast}
\end{array} \right)
\left(\begin{array}{c}
u_{{\bf j} \sigma} \\ v_{{\bf j} \bar{\sigma}}
\end{array} \right)
= E \left(
\begin{array}{c}
u_{{\bf i} \sigma} \\ v_{{\bf i} \bar{\sigma}}
\end{array} \right),
\end{equation}
where ${\bf i} = (x, y)$ is the site position in units of lattice spacing which is set to be $1$; $H_{{\bf ij}, \sigma} = -t \sum_{{\bf 1}} \delta_{{\bf i} + {\bf 1}, {\bf j}} + (\sigma h - \mu) \delta_{{\bf ij}} + U_0 \delta_{x,0}$; {\bf 1} denotes $(\pm 1, 0)$ and $(0, \pm 1)$; $\sigma = \pm 1$ is the spin index and $\bar{\sigma} = - \sigma$; $h$ is the Zeeman field; $\mu$ is the chemical potential; $\Delta_{{\bf ij}}$ is the OP and $\Delta_{{\bf ij}} = \Delta_{\bf i}\delta_{{\bf i},{\bf j}}$ for an $s$-wave superconductor; $u_{{\bf j} \sigma}$ and $v_{{\bf j} \bar{\sigma}}$ are the amplitudes of quasielectron and quasihole components, respectively. The proximity effect at the N/S junction interface is neglected and the OP of the S side is taken to be the bulk one since we are interested in the bulk properties here as in the original BTK theory. Here, we let the OP of the LO state be $\Delta_{{\bf i}} = \sum_{\alpha} \Delta_{\alpha} e^{i \alpha y}$, where the reciprocal lattice vector $\alpha = 0, 2\pi/a, \cdots, 2\pi (a - 1)/a$ and $a$ is the period of the OP. According to the Bloch theorem, the quasiparticle amplitude is a plane-wave factor with {\it crystal} $y$ momentum $K_y \in (-\pi/a, \pi/a]$ and true $x$ momentum $k_x \in (-\pi, \pi]$ times a function of period $a$ along $y$. For a given incident electron beam of spin $\sigma$ with energy $E$ and true $y$ momentum $k_y = K_y + \alpha$, we solve the BdG equations on a square lattice to obtain $B_{\alpha \alpha', \sigma}^{K_y}$ and $A_{\alpha \alpha', \bar{\sigma}}^{K_y}$, which are, respectively, the probabilities of reflected electrons and holes with true $y$ momentum $K_y + \alpha'$. The differential conductance of the N/S junction is then given by the Landauer-B\"uttiker-type formula
\begin{equation} \label{eq:Gns}
G_{\sigma}^{ns} = \frac{1}{L_y} \sum_{K_y, \alpha} \left[ 1 + \sum_{\alpha'} (A_{\alpha \alpha', \bar{\sigma}}^{K_y} - B_{\alpha \alpha', \sigma}^{K_y}) \right],
\end{equation}
where the junction size $L_y$ along $y$ is an integer multiple of $a$. It can be seen from Eq.~(\ref{eq:BdG}) that the dependence of $G_{\sigma}^{ns}$ on $\Delta E = E - \sigma h$ is the same for both $\sigma = \pm 1$ and the conductances due to $\sigma = +1$ and $-1$ incident electrons are not coupled. Thus, the total conductance $G^{ns}(E) = G_{+1}^{ns}(E) + G_{-1}^{ns}(E)$ and only the portion with one spin $\sigma$ will be considered in the following. Also, both the N/S conductance $G_{\sigma}^{ns}$ and the DOS in the superconducting state are divided by their corresponding normal-state values at each $E$ to yield the normalized $G_{\sigma}(E)$ and $\rho_{\sigma}(E)$, respectively.

\begin{figure*}
\includegraphics[width = 0.7\textwidth]{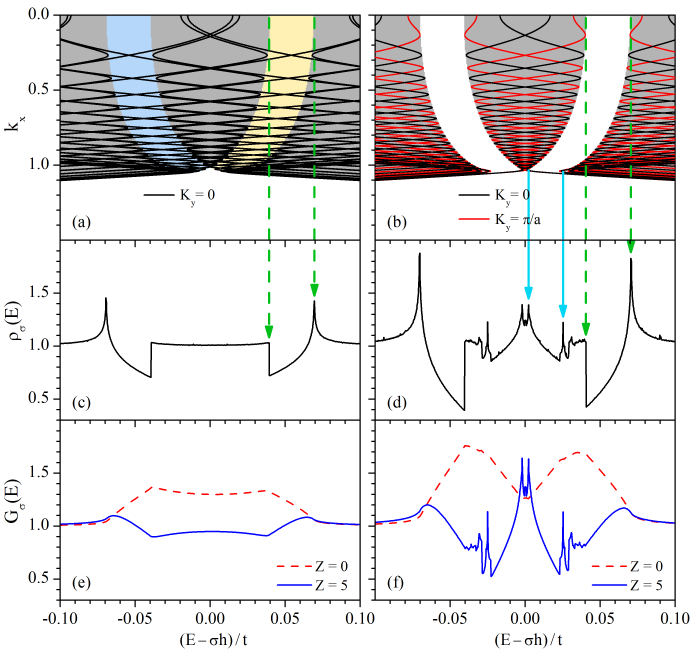}
\caption{\label{fig3} (Color online) Numerical simulations illustrating the situation with weak pairing interaction. The figures are the same as Figs.~\ref{fig2}(c)\,--\,\ref{fig2}(h) except $\Delta_q = 0.015$ and $a = 100$. The energy range is changed to $[-0.1,0.1]$ as a result of the small $\Delta_q$ compared with Fig.~\ref{fig2}.}
\end{figure*}

\begin{figure*}
\includegraphics[width = 0.7\textwidth]{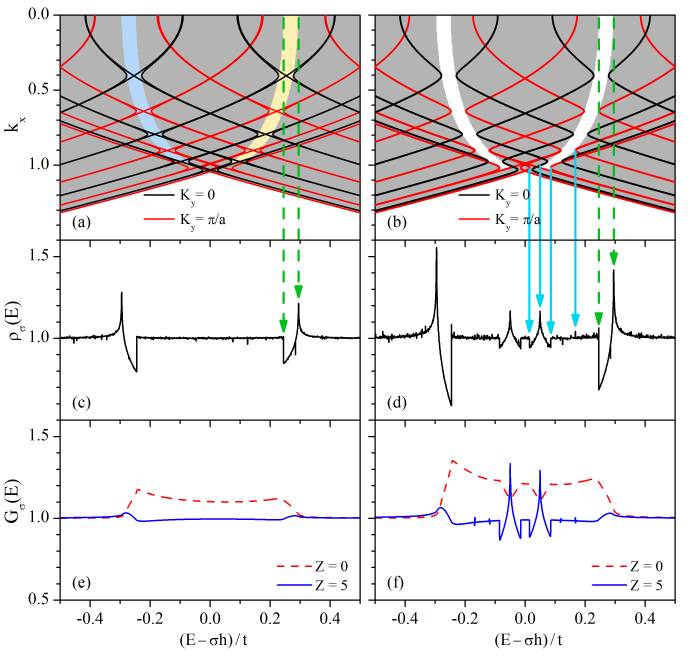}
\caption{\label{fig4} (Color online) Numerical simulations illustrating the situation with reduced Zeeman field $h/\Delta_{\mathrm{BCS}}$. The figures are the same as Figs.~\ref{fig2}(c)\,--\,\ref{fig2}(h) except $\Delta_q = 0.025$ and $a = 20$.}
\end{figure*}

In our numerical calculations, we take the chemical potentials of both sides to be the same $\mu = -3$, so that the Fermi surface is close to being circular and the Fermi momentum is roughly given by $k_F = 1$ for convenience. For a Zeeman field substantially above the lower critical field of the LO state, the OP can be approximated as $\Delta_{{\bf i}} = 2 \Delta_q \cos qy$, where $q = 2\pi/a$. We note that $a$ is determined by the strength of the pairing interaction (and thus the OP and the coherence length). However, for a given $h/\Delta_{\mathrm{BCS}}$, $\Delta_q a$ is fixed because it measures the ratio of the period length over coherence length [which is equivalent to $(q v_F / \Delta_q)^{-1}$ in a continuum model]. With increasing $h/\Delta_{\mathrm{BCS}}$, $\Delta_q a$ will decrease. A self-consistent calculation of the OP with $h/\Delta_{\mathrm{BCS}} = 0.8$ yields $\Delta_q a \approx 1.5$ in our model. Therefore, in Fig.~\ref{fig2}, we take $\Delta_q = 0.075$ and $a = 20$ as an example to illustrate the physics. We also calculate two more cases with $\Delta_q = 0.015$, $a = 100$ (Fig.~\ref{fig3}), and $\Delta_q = 0.025$, $a = 20$ (Fig.~\ref{fig4}) to show the situations with weak pairing interaction and strong Zeeman field, respectively. The junction size along $y$ is fixed at $L_y = 200\,000$ and periodic boundary condition is adopted. The $x$ momentum of the incident electron and those of the reflected electron and hole are not assumed to be equal in magnitude as in the BTK theory. Therefore, the corresponding group velocities are unequal, causing the $G_{\sigma}(E)$ spectra to be asymmetric about $\Delta E = 0$.

\section{Manifestations of hybridized midgap-states band in conductance spectra}

Before investigating the spectroscopic features of the LO and FF states, we need to first understand their band structures. For the FF state, due to the non-zero pairing momentum, the quasiparticle energy is given by
\begin{equation} \label{eq:FF}
\Delta E_{{\bf k}, {\bf q}} = \xi_{{\bf k}, {\bf q}}^{(a)} \pm \sqrt{\xi_{{\bf k}, {\bf q}}^{(s)2} + {\Delta_{q}}^2},
\end{equation}
where $\xi_{{\bf k}, {\bf q}}^{(a)} = (\xi_{{\bf k}+{\bf q}/2} - \xi_{-{\bf k}+{\bf q}/2})/2$, $\xi_{{\bf k}, {\bf q}}^{(s)} = (\xi_{{\bf k}+{\bf q}/2} + \xi_{-{\bf k}+{\bf q}/2})/2$, and $\xi_{{\bf k}} = -2t (\cos k_x + \cos k_y) - \mu$ is the kinetic energy of a $+{\bf k}$ electron, relative to $\mu$. From Eq.~(\ref{eq:FF}), we see that the gap of size $\Delta_q$ is no longer centered at $\Delta E = 0$, but is shifted by $\xi_{{\bf k}, {\bf q}}^{(a)}$, which has the same sign as $k_y$ for pairing momentum $+{\bf q}$ [Fig.~\ref{fig2}(b)]. Since each gap shift in the FF case involves a single sign of $k_y$, such that the dispersion curves for the opposite sign of $k_y$ cross inside the shifted gaps, we obtain a quasiparticle dispersion without a clear gap for the FF state [Fig.~\ref{fig2}(c)].~\cite{FFgap} For the LO state, however, there is now also pairing between $({\bf k}-{\bf q}/2, \uparrow)$ and $(-{\bf k}-{\bf q}/2, \downarrow)$, which causes a large number of $y$-momentum states to be coupled. In essence, our numerical treatment reveals that the crossings shown in the dispersion curves of the FF case become anticrossings. Consequently, a $k_x$-dependent gap appears in the dispersion, centered at $\Delta E = 0$, with a HMSB lying inside this gap [Figs.~\ref{fig2}(b) and \ref{fig2}(d)]. The surviving gaps on the two sides of the HMSB have roughly the same sizes and locations as the Zeeman-shifted gaps in the FF case because they actually arise from the crossing/anticrossing conversion.

As a result, bulk DOS with several singularities are obtained as shown in Figs.~\ref{fig2}(e) and \ref{fig2}(f). Here, we find two types of singularities. One type is related to states with momenta parallel to {\bf q} (i.e., with $k_x = 0$; see dashed arrows in Fig.~\ref{fig2}); the other type appears to be always associated with states with large $k_x$, and therefore with momenta nearly perpendicular to {\bf q} (thus along the nodal lines; see solid arrows of the same figure). The momentum directions of the states contributing to these singularities can be more easily understood by referring the values of $k_x$ on the Fermi surface shown in Fig.~\ref{fig2}(b). The singularities of each type can be further divided into two categories according to their origins, one due to the HMSB states and the other from the outer edges of the surviving gaps. For the ``parallel-momenta'' type of singularities, the energy difference of the singularities of the two origins is about $2\Delta_q$ and thus decreases with increasing Zeeman field [see Figs.~\ref{fig2}(f) and \ref{fig4}(d)]. For the ``perpendicular-momenta'' singularities, the ones due to the HMSB states approach $\Delta E = 0$ upon increasing the period length $a$ with fixed $h/\Delta_{\mathrm{BCS}}$ and there is always a gap separating the singularities of the two origins apart [see Figs.~\ref{fig2}(f) and \ref{fig3}(d)]. In the FF state, we only have the ``parallel-momenta'' singularities but not the ``perpendicular-momenta'' singularities because the HMSB is absent in the FF state.

In the point-contact or {\em metallic}-junction limit ($Z = 0$), Andreev reflection can occur with 100\% probability for $E$ inside a clean gap to enhance $G_{\sigma}(E)$ by a factor of exactly 2 since one hole retroreflected in N means one electron from within the Fermi sea of N has also moved from N into S. However, this enhancement would be reduced if transmission across the N/S interface could proceed via quasiparticle states at the incoming energy. These tendencies are well manifested in the $s$-wave BCS case, where $G_{\sigma}(E)$ is exactly $2$ inside the gap and reduces gradually to $1$ outside the gap by virtue of an energy-dependent transmission coefficient.~\cite{BTK} For the FF state, $G_{\sigma}(E)$ can not exceed $1.5$ since the $k_x$-dependent shift of the gap renders Andreev reflection nondominant in any energy range after summing over all $k_y$ states [Fig.~\ref{fig2}(g)]. For the LO state, in contrast, $G_{\sigma}(E)$ can exceed $1.5$ for energies within the surviving gaps, which occur on the two sides of the HMSB, where Andreev reflection can occur for a large range of $k_x$ [red dashed line in Fig.~\ref{fig2}(h)]. Here, $G_{\sigma}(E)$ can not reach $2.0$ for any energy since the HMSB essentially consists of quasiparticle states that can now facilitate the transmission of the incident electrons into S. This quasiparticle transmission, even if reduced from 100\% by N/S impedance mismatch, does diminish the subgap enhancement of $G_{\sigma}(E)$ and cause it to show a dip in an energy range around $\Delta E = 0$ where the HMSB exists.

\begin{figure}
\includegraphics[width = 0.39\textwidth]{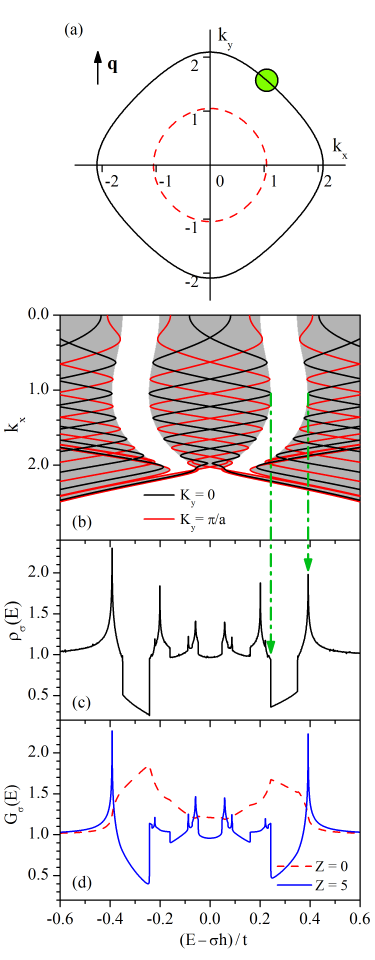}
\caption{\label{fig5} (Color online) Numerical simulations illustrating the situation with chemical potential $\mu$ increased for the LO state. Here, $\mu = -1$, $\Delta_q = 0.075$, and $a = 20$. (a) The Fermi surface of $\mu = -1$ (solid line) compared with that of $\mu = -3$ (dashed line). (b), (d) The same as Figs.~\ref{fig2}(d), \ref{fig2}(f), and \ref{fig2}(h). The singularities in DOS marked by green dashed-dotted arrows are related to states with momenta neither parallel nor perpendicular to {\bf q} [see the circled green shading in (a)].}
\end{figure}

As already discovered in our BTK model for the FF state,~\cite{FFcase} the {\em tunneling} ($Z \gg 1$) conductance measures the weighted DOS that sums over states with a projection factor, which favors quasiparticle states with momenta nearly perpendicular to the N/S interface when only the bulk properties are involved. Therefore, the ``parallel-momenta'' singularities shown in the bulk DOS are suppressed in the tunneling conductance spectrum [Fig.~\ref{fig2}(h)] when the N/S interface is parallel to {\bf q}. However, these ``parallel-momenta'' singularities are expected to reemerge when their corresponding momenta are no longer parallel to the interface, such as in the situation where the interface is tilted from the present direction. An interesting situation to illustrate this expectation can be achieved by hypothetically raising the chemical potential to $\mu = -1$ so that the Fermi surface is changed from a nearly circular shape to a squarish shape [Fig.~\ref{fig5}(a)]. Here, we obtain two singularities associated with momenta neither parallel nor perpendicular to {\bf q} [see dashed-dotted arrows in Fig.~\ref{fig5} and the circled shading in Fig.~\ref{fig5}(a)], and these singularities appear in the tunneling conductance as expected. As to the ``perpendicular-momenta'' singularities, they are faithfully reproduced in the tunneling conductance spectrum of our choice of the interface orientation because they are essentially unaffected by the projection-factor weighting. For the FF state, the ``parallel-momenta'' singularities [Fig.~\ref{fig2}(e)] show very similar behavior [Fig.~\ref{fig2}(g)].

These observations agree well with the HMSB physics reviewed in the Introduction. For the LO state, the HMSB states, formed around the nodal lines, produce a broad DOS hump near $\Delta E = 0$ as well as ``perpendicular-momenta'' singularities within this hump due to accumulation of spectral contributions. When the interface intercepts the nodal lines, the nodal lines of the OP behave effectively as channels for transmission because of the hybridized midgap states formed mainly near and along these nodal lines. Thus, electrons can be transmitted from N to S as HMSB quasiparticles, causing a decrease in the metallic-junction conductance due to reduced probability for Andreev reflection, and an increase in the tunneling conductance due to resonant tunneling. This junction orientation can always be realized on a finite-sized sample. Note that, provided the size of the superconducting sample is not large enough to accommodate multidomains of the LO state, we have only one direction of the nodal lines for the whole sample. It is then possible to pick one of the several differently oriented faces of the sample for making junctions so that the nodal lines intercept the interface, and the manifestations of the HMSB discussed here can be observed. These manifestations of the HMSB can also survive in the presence of disorder due to the topological origin of the midgap states. As shown in Ref.~\onlinecite{loh}, when the disorder strength is not strong enough to destroy the LO state, a broad hump of lower height appears inside the gap and around $\Delta E = 0$ in the DOS spectrum as a result of the HMSB. However, the DOS singularities discussed here are eliminated because disorder will spread the energies of the originally accumulated states to a larger energy range. We therefore find that the manifestations of the HMSB in the junction conductance, which are consequences of the unique periodically sign-reversing structure of the LO OP, can be used to identify the LO state effectively.

\section{Conclusion}

In conclusion, by applying a discrete square-lattice BTK model to a spatially periodic superconducting OP, we have investigated signatures of the FFLO state in the N/S conductance spectroscopy. We have focused on the $s$-wave LO case with the N/S interface oriented along the pairing momentum {\bf q}, expanding on previous works, which include our prior treatment of the FF case. Unique to the LO case is the HMSB, which is formed amid a momentum-dependent gap as a result of the periodic OP sign reversal. These HMSB states are hybridized from essentially dispersionless midgap quasiparticle states localized along the nodal lines and can help the transmission of incident electrons into the superconductor. This specific band-in-gap structure is thus shown to give rise to distinctive conductance features, which are absent in the FF case. Our results are generically robust, i.e., they are expected to be qualitatively valid for all junction orientations where the nodal lines are intercepted by the N/S interface, and in the presence of disorder as long as the LO state is not destroyed. We therefore conclude that these generic manifestations of the HMSB and the surviving gaps discussed here can be systematically probed with tunneling and point-contact spectroscopy on oriented sample surfaces, to provide clear experimental signatures for distinguishing the LO state from the FF state, and both from the BCS state. 

\acknowledgments

This work was supported by: NSFC Grant No.~11047136, Yunnan Provincial Applied Basic Research Foundation (2009CD094), Qujing Normal University (2008ZD005), and DOE Grant No.~DE-FG52-10NA29659 (Q.C.); National Science Foundation Grants No.~DMR-0704133 and No.~DMR-1004545 (K.Y.); NSERC, CFI/OIT, and Canadian Institute for Advanced Research under the Quantum Materials Program (J.Y.T.W.). Q.C. is grateful to X. Wan for hospitality at Asia Pacific Center for Theoretical Physics and Z.-X. Hu for computer help.


\begin{thebibliography}{99}
\bibitem{transmission}
Transmission perpendicular to (or across) the nodal lines is expected to be weaker since it must depend on the weak overlap of the midgap-states wavefunctions in-between the nodal lines. This will be particularly true at lower Zeeman fields when the nodal lines are farther apart.
\end{thebibliography}
\end{document}